# Manipulating Photogalvanic Effects in Two-Dimensional Multiferroic Breathing Kagome Materials

Haonan Wang[1] and Li Yang[1,2*]

1. Department of Physics, Washington University, St. Louis, MO, 63130, USA

2. Institute of Materials Science and Engineering, Washington University, St. Louis, MO, 63130, USA

*lyang@physics.wustl.edu

**Abstract**

Multiferroic materials, known for their multiple tunable orders, present an exceptional opportunity to manipulate nonlinear optical responses, which are sensitive to symmetry. In this study, we propose leveraging electric and magnetic fields to selectively control and switch specific types of photogalvanic effects in two-dimensional multiferroic breathing kagome materials. Taking monolayer $Nb_3I_8$ as an example, we demonstrate that the shift current, characterized by the real-space shift of electrons and holes, is predominantly unaffected by magnetic order. In contrast, injection current, featured by quantum metric dipole in momentum space, is closely related to valley polarization which can be controlled by magnetic field. Furthermore, both photocurrents can be reversed by out-of-plane electric field via the lattice breathing. Our findings reveal the potential of multiferroic beathing kagome structures for multifunctional optoelectronic applications and sensors.

Keywords: breathing kagome lattice, photogalvanic effect, multiferroic material, optoelectronic manipulation, first-principles calculations

**Introduction**

Under optical irradiation, insulating materials with inversion symmetry breaking [1–3] can generate a DC photocurrent via second-order light-matter interactions. This includes the presence of shift current [4–7] and injection current [8–10], commonly referred to as the photogalvanic effect or bulk photovoltaic effect. These two phenomena originate from the real-space displacement of wave packets and the velocity injection of electrons and holes, respectively [2,11,12]. Recently, numerous novel photogalvanic processes have emerged in quantum materials with specific symmetries, such as the magnetic bulk photovoltaic effect [9,10,13–15], nonlinear anomalous Hall effect [16], quantized circular photogalvanic effect [17], and spin photocurrent [18–23]. Most of these effects have been observed in parity-time-symmetric (PT) antiferromagnetic insulators [9,10,14,15,19] as well as non-centrosymmetric ferroelectric semiconductors [16,20,23] and Weyl semimetals [17,24–30]. These novel photogalvanic effects are deeply rooted in the quantum geometry of electron wavefunctions, such as quantum metric and Berry curvature [24,31,32], providing a promising tool to detect emerging quantum phases.

In addition to generating various photogalvanic effects, tuning them in a controllable way is also highly desirable. Multiferroic materials appear to be ideal candidates for such manipulations, given their switchable multiferroic orders interconnected via specific symmetries. For instance, recent proposals have been made to manipulate injection current in stacking-induced multiferroic bilayers [13]. However, the manipulation of shift current is constrained by symmetry limitations. Therefore, it is crucial to explore switchable injection and shift currents in two-dimensional (2D) multiferroic materials, despite only a limited number of 2D multiferroic materials having been predicted over the past decade [33–37].

In this work, we propose to selectively manipulate both injection and shift currents in a family of 2D breathing kagome materials: $Ta_3X_8$ (X=Br, I) and $Nb_3X_8$ (X=Cl, Br, I). These materials have aroused significant research interest as a promising platform for topological flat bands [38], intrinsic anomalous valley Hall effect [39], Mott insulator [40], and multiferroics [41–44]. Using first-principles calculations, we demonstrate that the unique lattice breathing allows for the selective tuning and switching of injection and shift currents through electric and magnetic fields. For instance, in monolayer $Nb_3I_8$, we find that the valley polarization, which causes asymmetric distribution of quantum metric dipole in momentum space, can couple with both the out-of-plane ferromagnetic (FM) order and ferroelectric (FE) order. Consequently, when the magnetic order or electric polarization is reversed, the direction of the injection current also reverses. Conversely, shift current only depends on the in-plane lattice breathing instead of valley polarization. As a result, we predict the presence of four distinct nonlinear optical responses corresponding to different multiferroic configurations, providing a method for selectively manipulating nonlinear photocurrents in 2D breathing kagome lattice.

**Results and Discussion**

The general form of second-order DC photocurrent can be written as

$$J^a = \chi^a_{bc}(0; \omega, -\omega) E_b(\omega) E_c(-\omega), \quad (1)$$

where the subscripts $b$ and $c$ denote the polarization directions of electric field ($E$) of incident light, and the superscript $a$ refers to the photocurrent direction in Cartesian coordinates. For insulators, the photoresponsivity $\chi^a_{bc}(0; \omega, -\omega)$ is mainly contributed from interband excitations, including shift current and injection current [2]. Under linearly polarized light, through the quantum perturbation theory within the independent particle approximation, the photoconductivity tensors of shift ($\sigma_{bc}^a$) and injection ($\eta_{bc}^a$) currents in the length gauge can be expressed as [2,9,10]

$$\sigma_{bc}^a(0;\omega,-\omega) = -\frac{i\pi e^3}{\hbar^2}\sum_{m,n}\int\frac{d\mathbf{k}}{(2\pi)^2}\{r_{mn}^b, r_{nm}^{c;a}\}\delta(\omega-\omega_{mn}), \quad (2)$$

$$\eta_{bc}^a(0;\omega,-\omega) = -\frac{\pi e^3}{\hbar^2}\sum_{m,n}\int\frac{d\mathbf{k}}{(2\pi)^2}\Delta_{mn}^a\{r_{mn}^b, r_{mn}^c\}\tau\delta(\omega-\omega_{mn}), \quad (3)$$

where $r_{mn}^b = \frac{v_{mn}^b}{i\omega_{mn}}$ is the interband dipole matrix element between the conduction band $m$ and valence band $n$. $\{r_{mn}^b, r_{nm}^{c;a}\}$ in Eq. (2) is defined as $(r_{mn}^b r_{nm}^{c;a} + r_{mn}^c r_{nm}^{b;a})$. $r_{nm}^{c;a} = \frac{\partial r_{nm}^c}{\partial k^a} - i(\mathcal{A}_{nn}^a - \mathcal{A}_{mm}^a)r_{nm}^c$ is the generalized derivative with respect to the crystal momentum $k$ where $\mathcal{A}_{mm}^a$ is the Berry connection matrix element. $\{r_{mn}^b, r_{mn}^c\}$ is the Hermitian metric tensor defined as $r_{mn}^b r_{nm}^c + r_{nm}^b r_{nm}^c$. Such Hermitian metric is also known as quantum geometric tensor and is distinct from the Fubini-Study metric in the nonlinear Hall effect [45]. $\Delta_{mn}^a = v_{mm}^a - v_{nn}^a$ denotes the group velocity difference where $v_{mm}^a$ stands for the velocity matrix element. The integrand of injection current $\Delta_{mn}^a\{r_{mn}^b, r_{mn}^c\}$ in Eq. (3) is termed as quantum metric dipole [46]. $\tau$ denotes the carrier lifetime and is set to 0.1 ps in our calculations (see SI). Both shift current and injection current are rooted in quantum geometry of electron wave functions [24,31].

It is noteworthy that injection current and shift current have essentially distinct parities under spatial inversion and time reversal operations [24]. Under time-reversal transformation $\mathcal{T}$, $\mathcal{T}r_{mn}^a(\mathbf{k}) = r_{mn}^{*a}(-\mathbf{k}) = r_{nm}^a(-\mathbf{k})$, and the Berry connection is even, namely, $\mathcal{T}\mathcal{A}_{mm}^b(\mathbf{k}) = \mathcal{A}_{mm}^b(-\mathbf{k})$. Hence, both terms $\{r_{mn}^b, r_{mn}^c\}$ and $\{r_{mn}^b, r_{nm}^{c;a}\}$ are even in the momentum space. Therefore, according to Eq. (2), shift current is immune to time-reversal operation $\mathcal{T}$. On the other hand, the group velocity difference $\Delta_{mn}^a$ in Eq. (3) is odd in momentum space because $\mathcal{T}v_{mm}^a(\mathbf{k}) = -v_{mm}^{*a}(-\mathbf{k}) = -v_{mm}^a(-\mathbf{k})$. Thus, injection current is odd under $\mathcal{T}$. Under spatial inversion transformation $\mathcal{P}$, denoted by $\mathcal{P}v_{mn}^a(\mathbf{k}) = -v_{mn}^a(-\mathbf{k})$, $\mathcal{P}r_{mn}^a(\mathbf{k}) = -r_{mn}^a(-\mathbf{k})$, and $\mathcal{P}r_{nm}^{b;a}(\mathbf{k}) = r_{nm}^{b;a}(-\mathbf{k})$. Consequently, both injection and shift currents reverse their directions by the inversion operation $\mathcal{P}$. Based on the above symmetry analysis, the spatial inversion associated with FE polarization switching and time reversal associated with magnetic order switching can be employed to selectively control injection and shift currents, respectively, in multiferroic materials.

*Monolayer kagome Nb₃I₈:* To demonstrate this idea, a material with both controllable parity and time reversal symmetries is required, bringing our attention to a family of emerging 2D kagome materials, niobium halide semiconductors, which have attracted significant research interest because of their unique multiferroic characters and topological properties. Here we choose monolayer $Nb_3I_8$ as an example, which has a space group of $P3m1$. As shown in Figure. 1 (a) and (b), the atomic structure has a three-fold in-plane rotational symmetry ($C_{3v}$) and a mirror symmetry with respect to the $yz$ mirror plane ($M_x$). Three Nb ions are trimerized as a $Nb_3$ cluster in the unit cell. Two types of $Nb_3$ trimers with different bond lengths form a distorted Kagome lattice that breaks the inversion symmetry, leading to an out-of-plane polarization [41,42]. Interestingly, the out-of-plane polarization is inherently coupled with the lattice breathing. As a result, FE polarization switching by a vertical electric field is also accompanied by the in-plane inversion of atomic configuration. Finally, 1 $\mu_B$ magnetic moment for each $Nb_3$ trimer has been calculated, consistent with previous results [39,47]. Hence, the FE/FM orders in monolayer $Nb_3I_8$ can be switched via $\mathcal{P}/\mathcal{T}$ symmetry operation.

The electronic band structures of monolayer $Nb_3I_8$ are plotted in Figure. 1 (c) and (d). When the spin orientation is along the in-plane $x$-axis (**M** ∥ **x**), the energy bands are symmetric in Figure. 1 (c). When the spin orientation aligns along the out-of-plane $z$-axis (**M** ∥ **z**), the band structure presents an energy asymmetry at **k** and -**k**, resulting in a valley polarization. As shown in circled areas, the energy difference between for the bottoms of the second conduction band at $\mathbf{k} = \mathbf{K}$ and $\mathbf{k} = -\mathbf{K}$ is around 80 meV, which can quantitatively describe the valley polarization. Notably, the valley polarization is reversed between $\mathbf{k}$ and $-\mathbf{k}$ points when the out-of-plane spin orientation is flipped ($\mathbf{M} \parallel -\mathbf{z}$) in Figure. 1(d) because the spin flipping is achieved through time reversal.

*Linear optical response:* Figure. 2(a) exhibits a broad linear optical response profile. Optical transitions between the top valance band and bottom conduction band are forbidden because of their opposite spins [39], and the allowed lowest-energy transition occurs between the top valance band and the second lowest-energy conduction band around the valleys, $\mathbf{k} = \mathbf{K}$ and $\mathbf{k} = -\mathbf{K}$, at an energy of 0.6 eV. The linear optical spectra (the imaginary part of $\varepsilon_{yy}$) with different spin orientations ($\mathbf{M} \parallel \mathbf{x}$, $\mathbf{M} \parallel \mathbf{y}$, and $\mathbf{M} \parallel \mathbf{z}$) are also plotted, which are not sensitive to the spin orientation. Moreover, we also find that they are degenerate for two FE phases because the linear response is even under $\mathcal{P}$ [2].

*Shift current:* Because of the second-order nature of the photogalvanic effect, it is crucial to identify the non-zero independent components of the photoconductivity tensor. We note that in our investigations presented below, we only focus on the photo-induced charge current with in-plane components. The photoconductivities are calculated by equation (2) and (3) via summation across all bands regardless of spins. As analyzed above, shift current is immune to time reversal. Therefore, its independent components are only constrained by the space group symmetry, i.e., the mirror symmetry $M_x$ and rotational symmetry $C_{3v}$. Because the dipole matrix element $r_{mn}^a$ is a polar vector in momentum space, $r_{mn}^x$ is odd and $r_{mn}^y$ is even with respect to the $M_x$ mirror plane. In consequence, all the nonvanishing tensor components are constrained to $\sigma_{yy}^y = -\sigma_{xx}^y = -\sigma_{xy}^x = -\sigma_{yx}^x$, and we only analyze one component, e.g., $\sigma_{yy}^y$, in Figure. 2(b).

Firstly, significant shift current is observed. The magnitude of $\sigma_{yy}^y$ reaches 22 $\mu A/V^2$. This is larger than that (~8 $\mu A/V^2$) of monolayer MoS$_2$ [48] and closed to that (~28 $\mu A/V^2$) of sliding-induced multiferroic bilayer VS$_2$ [13]. Secondly, as expected, the shift current is not sensitive to the spin orientation. In Figure. 2(b), the shift current spectrum is nearly identical with different spin orientations (**M** ∥ **x**, **M** ∥ **y**, and **M** ∥ **z**), where the minor changes are owing to the variation of band structures.

Furthermore, we have calculated shift current with respect to the polarization direction of incident light, which can be directly measured in experiment. We only consider the direction of the incident light along the out-of-plane direction in the whole work. This is because the depolarization effect via the electron-hole exchange interaction will quench the out-of-plane electric field of light and subsequent optical response [49]. For light with an in-plane polarization direction, its optical field is $|E_0|e^{-i\omega t}(\cos\phi, \sin\phi) + c.c.$. The in-plane shift current can be calculated as $(J^x, J^y) = |E_0|\sigma_{yy}^y(-\sin 2\phi, \cos 2\phi)$ and is plotted in Figures. 2 (c) and (d), respectively. In Figure. 2 (c), the shift current along the $x$ direction ($J^x$) reaches its maximum amplitude when the light is polarized at the 45-degree angle to the current direction. On the other hand, for shift current along the $y$ direction ($J^y$) plotted in Figure. 2 (d), it attains the peak value when the light is polarized either along or perpendicular to the current direction. Finally, in line with the findings in Figure. 2 (b), all these shift currents exhibit low sensitivity to the spin orientation.

*Injection current:* Unlike shift current, injection current is highly dependent on the spin orientation. For **M** ∥ **x**, the symmetry operation is $M_x$. According to Eq. (3), because the velocity matrix element $v_{mm}^x$ behaves as a polar vector in momentum space, $\Delta_{mm}^x$ is odd and $\Delta_{mm}^y$ is even with respect to the $yz$ ($M_x$) mirror plane. Hence there are three independent non-zero components,

$\eta_{yy}^{y}$, $\eta_{xx}^{y}$, and $\eta_{xy}^{x} = \eta_{yx}^{x}$. For **M** ∥ **y**, the symmetry operation is $M_x\mathcal{T}$. $\{r_{mn}^b, r_{mn}^c\}$ is even for $b = c$, and odd for $b \neq c$ under the combined $M_x\mathcal{T}$ symmetry. Meanwhile, the group velocity difference $\Delta_{mn}^x$ is even, and $\Delta_{mn}^y$ is odd under $M_x\mathcal{T}$. Therefore, there are three independent components, $\eta_{xx}^x$, $\eta_{yy}^x$, and $\eta_{xy}^y = \eta_{yx}^y$. For **M** ∥ **z**, the symmetry operations are $M_x\mathcal{T}$ and $C_{3v}$ that enforce only one independent component to survive, $\eta_{xx}^x = -\eta_{yy}^x = -\eta_{xy}^y = -\eta_{yx}^y$. In the following, we focus on the $\eta_{yy}^x$ component as an example. The results of other components are presented in the supplementary information.

Figure 3 (a) plots the spectra of $\eta_{yy}^x$ component with different spin orientations. Overall, an enhanced response of injection current is observed. For **M** ∥ **z**, the first peak at an incident light energy of 0.6 eV reaches ~100 $\mu A/V^2$, larger than that in sliding-induced multiferroic VS$_2$ bilayer (~10 $\mu A/V^2$) [13] and MoS$_2$ monolayer (~$10^{-6}$ $\mu A/V^2$) [50].

Importantly, the injection current is highly sensitive to spin orientations. We focus on the absorption edge region, which is the shadow area in Figure. 3(a). The first peak of $\eta_{yy}^x$ reaches its maximum (~ 100 $\mu A/V^2$) when **M** ∥ **z**. Then it begins to decrease as the spin orientation rotates into the in-plane direction. When the spin orientation is along the in-plane $y$ direction, the magnitude decreases to 12 $\mu A/V^2$. Finally, when **M** ∥ −**z**, the magnitude of $\eta_{yy}^x$ reaches the maximum again but with an opposite sign. We note that $\eta_{yy}^x$ becomes zero when **M** ∥ **x** because of the symmetry constrain.

Based on Eq. (1) and symmetries of the injection current tensors, the in-plane injection current components $J^x$ and $J^y$ are plotted in Figures. 3(b) and (c), respectively. Consisting with the result of the $\eta_{yy}^x$ component, the injection current is significantly larger for the out-of-plane spin orientation, and substantially reduced for the in-plane spin orientation. In the case of **M** ∥ **z**, the injection current along the *x* direction ($J^x$) reaches its maximum amplitude when the incident light is polarized either along or perpendicular to the current direction. While for the injection current along the *y* direction ($J^y$) plotted in Figure. 3 (c), it attains the peak value when the incident light is polarized at an angle of 45 degrees to the current direction. Compared with the results of shift current in Figures. 2 (c) and (d), the angle dependence of both current components is inversed in injection current. It is noteworthy that when the in-plane incident light is parallel or perpendicular to the spin orientation, the in-plane injection current is larger in directions perpendicular to the in-plane spin orientation, namely, $J^x$ for **M** ∥ **y** is larger than that for **M** ∥ **x**, and vice versa for $J^y$, this phenomenon has also been observed in previous research [13].

To better address this characteristic of injection current, we have plotted the variation of the peak value of injection current with respect to spin orientation in Figure. 3 (d). It is clearly seen that both injection current components significantly vary with the spin orientation and can switch the direction by flipping spins. Moreover, we find that the amplitude and direction of injection current is strongly correlated to the valley polarization. A large valley polarization accompanies with a large injection current, indicating that the valley-polarization induced asymmetric band structure (Figure. 1 (d)) plays an important role in determining the amplitude of injection current.

To elucidate the underlying mechanism of spin dependence, we analyze the distribution of group velocity difference and quantum metric dipole in momentum space at the photon energy of 0.6 eV for **M** ∥ **y** and **M** ∥ **z**, corresponding to the interband transitions around **k** = **K** and **k** = −**K** in reciprocal space, as marked by the circles in Figure. 1 (d). According to Eq. (3), the injection current photoconductivity $\eta^x_{yy}$ is the integral of quantum metric dipole that is the absorption rate $|r^y_{mn}|^2 \delta(\omega - \omega_{mn})$ weighted by group velocity difference $\Delta^x_{mn}$. Firstly, although the distribution of the absorption rate $|r_{mn}^{y}|^2 \delta(\omega - \omega_{mn})$ for **M** ∥ **y** is more symmetric than that for **M** ∥ **z** around **k** = **K** and **k** = −**K** points (see supplementary information), their integral that gives rise to the linear optical response (the imaginary part of $\varepsilon_{yy}$) is nearly identical, as shown in Figure. 2(a). Hence, the main difference of the nonlinear response for **M** ∥ **y** and **M** ∥ **z** comes from the group velocity difference $\Delta_{mn}^{x}$. The distributions of group velocity difference $\Delta_{mn}^{x}$ between the top valance band and bottom conduction band are plotted as Figures. 4(a) and (b) for **M** ∥ **y** and **M** ∥ **z**, respectively. Since both magnetic configurations preserve the $M_x \mathcal{T}$ symmetry, $\Delta_{mn}^{x}$ is symmetric with respect to the $xz$ mirror plane ($M_y$), while it is asymmetric with respect to the $yz$ mirror plane ($M_x$). Therefore, the quantum metric dipole $\Delta_{mn}^{x} |r_{mn}^{y}|^2$ is nonvanishing that gives rise to a net current of $\eta_{yy}^{x}$.

Moreover, for **M** ∥ **z** plotted in Figure. 4(b), $\Delta_{mn}^{x}$ exhibits greater asymmetry at **k** = **K** and **k** = −**K** points. This agrees with its larger valley polarization, which induces energy and group velocity asymmetry in momentum space (see Figure. 1(d)) and thereby enhances current injection. Conversely, in the case of in-plane magnetic orders, the valley polarization diminishes, allowing energy and group velocity symmetry to persist and suppressing the current injection. In consequence, the quantum metric dipole exhibits a more asymmetric distribution when **M** ∥ **z**, as illustrated in Figures. 4 (c) and (d), contributing to a larger integral value. This analysis demonstrates that the injection current originates from the asymmetry of quantum metric dipole in momentum space and strongly indicate that the injection current can be amplified by valley polarization. The fact that the injection current is enhanced via valley polarization when the magnetization vector is aligned along the out-of-plane direction has also been observed in MnPSe$_3$ monolayer [15].

*Selectively tunable nonlinear photocurrent:* Table I summarizes the properties of shift and injection currents in multiferroic 2D kagome materials and compares them with those of sliding multiferroic bilayers [13]. In sliding multiferroic bilayers, though injection current can be reversed by FE or FM switching, the shift current remains unaffected because of the protection of the horizontal mirror symmetry. On the other hand, in 2D breathing kagome lattices, the two FE phases are interrelated through $\mathcal{P}$, that can reverse both the injection and shift currents [24]. Furthermore, reversing the spin orientation in 2D kagome lattices leads to a reversal of the injection current, while the shift current remains unaffected.

This offers a unique tuning knob for separately controlling shift and injection photocurrents. Figure 5 illustrates the four possible multiferroic phases: **P↑M↑**, **P↑M↓**, **P↓M↓**, and **P↓M↑**, where the up/down arrow represents the out-of-plane direction of FE or FM order. The injection current can be solely reversed by switching the magnetic field, while the shift current can be solely reversed by switching both electric and magnetic fields. By applying an appropriate external field, we can selectively switch the specific type of DC photocurrent. Additionally, due to the different spectra and amplitudes of shift and injection currents, the four multiferroic phases exhibit different overall photocurrents, resulting in four-stage photocurrent states. This provides a novel approach to identifying multiferroic phases and can also serve as multifunctional sensors for detecting both electric and magnetic fields.

Finally, the conclusion about tunable photocurrents, as presented above, remains valid for diverse conditions, including those induced by circularly polarized light and varying Hubbard values, as detailed in supplementary information. Therefore, our approach to manipulate photocurrent inherently broadens its applicability across a wide range of multiferroic breathing kagome materials. Notably, one such breathing kagome material, $Nb_3Cl_8$, has been identified as a Mott insulator [40,51]. Investigations of photoinduced effect for Mott insulators go beyond the band theory extend beyond conventional band theory, necessitating the application of many-body theories such as Floquet dynamical mean-field theory [52,53]. We also anticipate future research to explore switchable photogalvanic effects a broader range of multiferroic materials.

**Supporting Information**

Computational details; k-point mesh convergence test; analyses of the photoconductivities of injection current in the monolayer $Nb_3I_8$; analyses of absorption rates of the monolayer $Nb_3I_8$; hybrid functional calculations of band structures and optical responses for monolayer $Nb_3I_8$; band structures and optical responses for U=3.0 eV; analyses of shift and injection current under circularly polarized light; second harmonic generation for monolayer $Nb_3I_8$.

**Acknowledgment**

H.W. is supported by National Science Foundation (NSF) DMR-2124934. L.Y. is supported by Designing Materials to Revolutionize and Engineer our Future (DMREF) DMR-2118779. The simulation used Anvil at Purdue University through allocation DMR100005 from the Advanced Cyberinfrastructure Coordination Ecosystem: Services & Support (ACCESS) program, which is supported by National Science Foundation grants #2138259, #2138286, #2138307, #2137603, and #2138296.

**Figures**

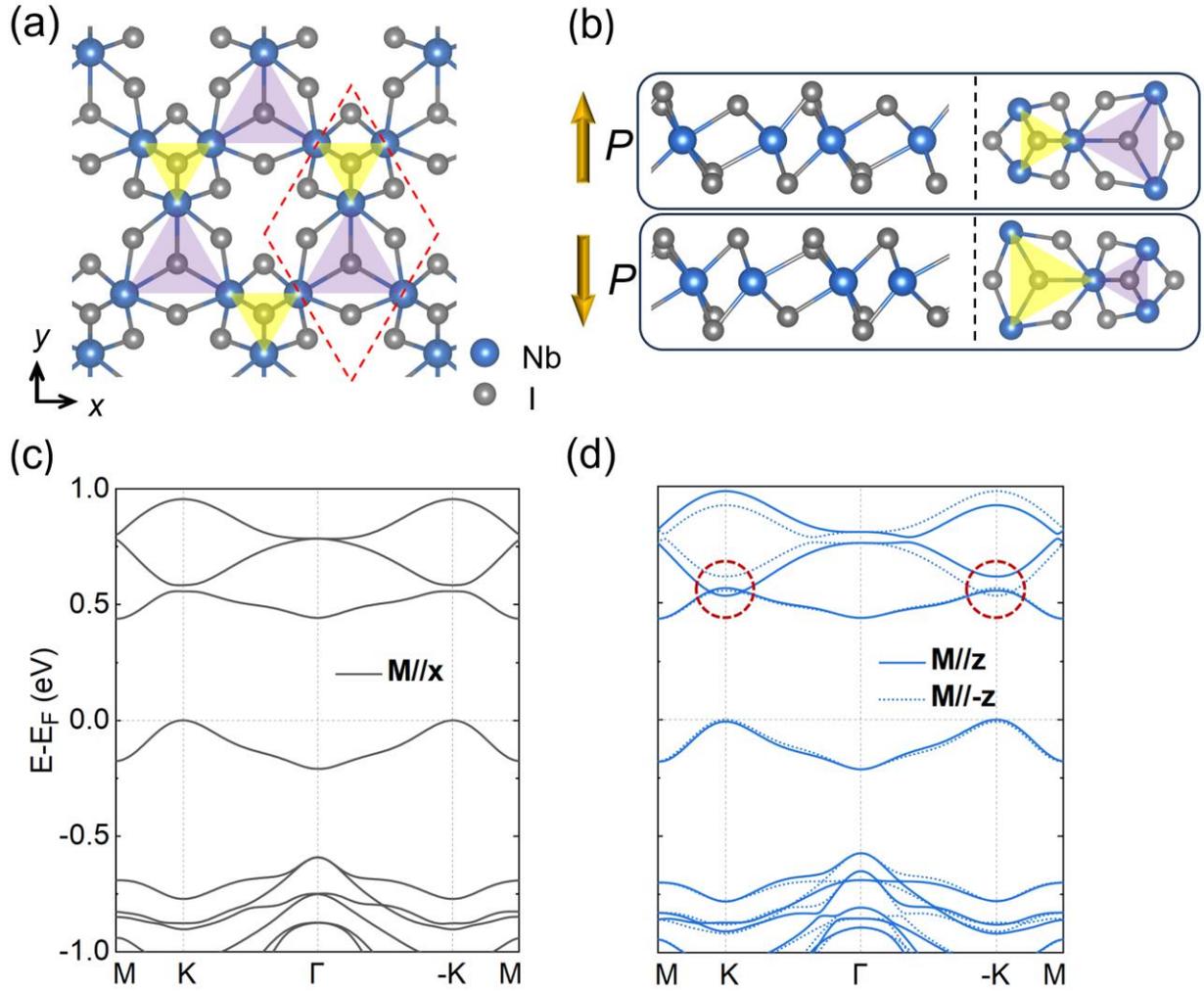

**Figure 1** (a)-(b) Top view and side view of atomic structure of the breathing kagome lattice Nb$_3$I$_8$ monolayer. The purple and yellow shaded regions represent two kinds of Nb trimers with different bond length. The two ferroelectric phases with opposite out-of-plane polarization can be switched through a lattice breathing that space inversion. (c) Band structure of Nb$_3$I$_8$ monolayer with $M \parallel x$, (d) Band structures for $M \parallel z$ (blue solid line) and $M \parallel -z$ (blue dotted line). Energy bands at $k$ and $-k$ are symmetrical for $M \parallel x$, while they are asymmetrical for $M \parallel z$, referred to as valley polarization. Bands in the dashed circles contribute to the first peak of optical interband transitions. All bands are reversed between $k$ and $-k$ for $M \parallel -z$ due to time reversal.

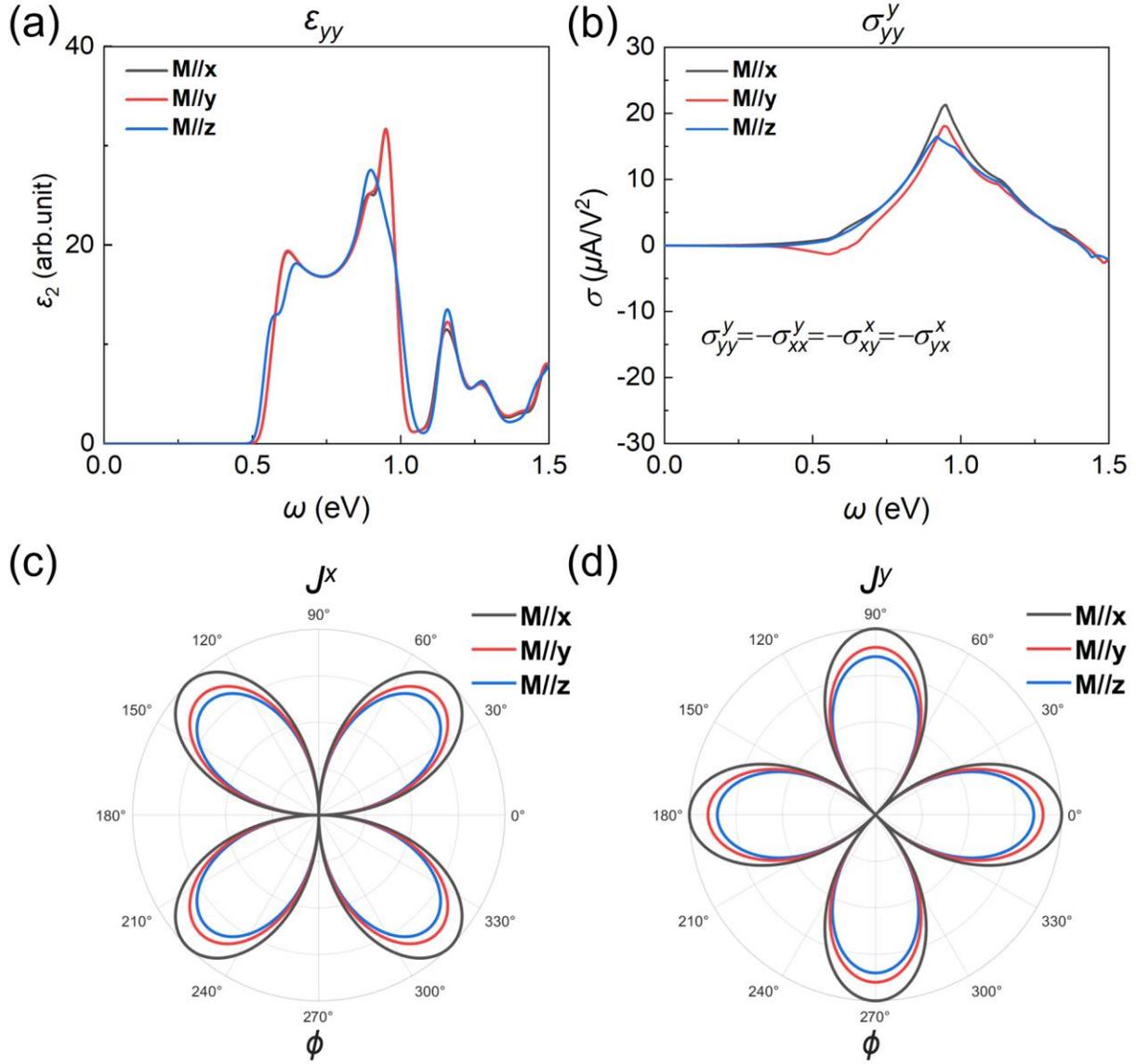

**Figure 2** Optical responses in Nb$_3$I$_8$ monolayer with varying magnetic orders. (a) Imaginary part of linear optical responses $\epsilon_{yy}$. (b) Shift current photoconductivity of $\sigma^{y}_{yy}$ component, other nonvanishing components reverse sign due to the three-fold rotational symmetry. All varying magnetic orders have same nonvanishing components. (c)-(d) Angle-dependent shift current $(J^x, J^y)$ at photon energy of 0.9 eV. Angular coordinate $\phi$ denotes the angle between the *x*-axis and incident light polarization direction, and radial coordinate stands for amplitude of photocurrents. The maximum amplitude of shift current $(J^x, J^y)$ for ***M*** ∥ ***z*** is comparable to that for ***M*** ∥ ***x*** and ***M*** ∥ ***y***.

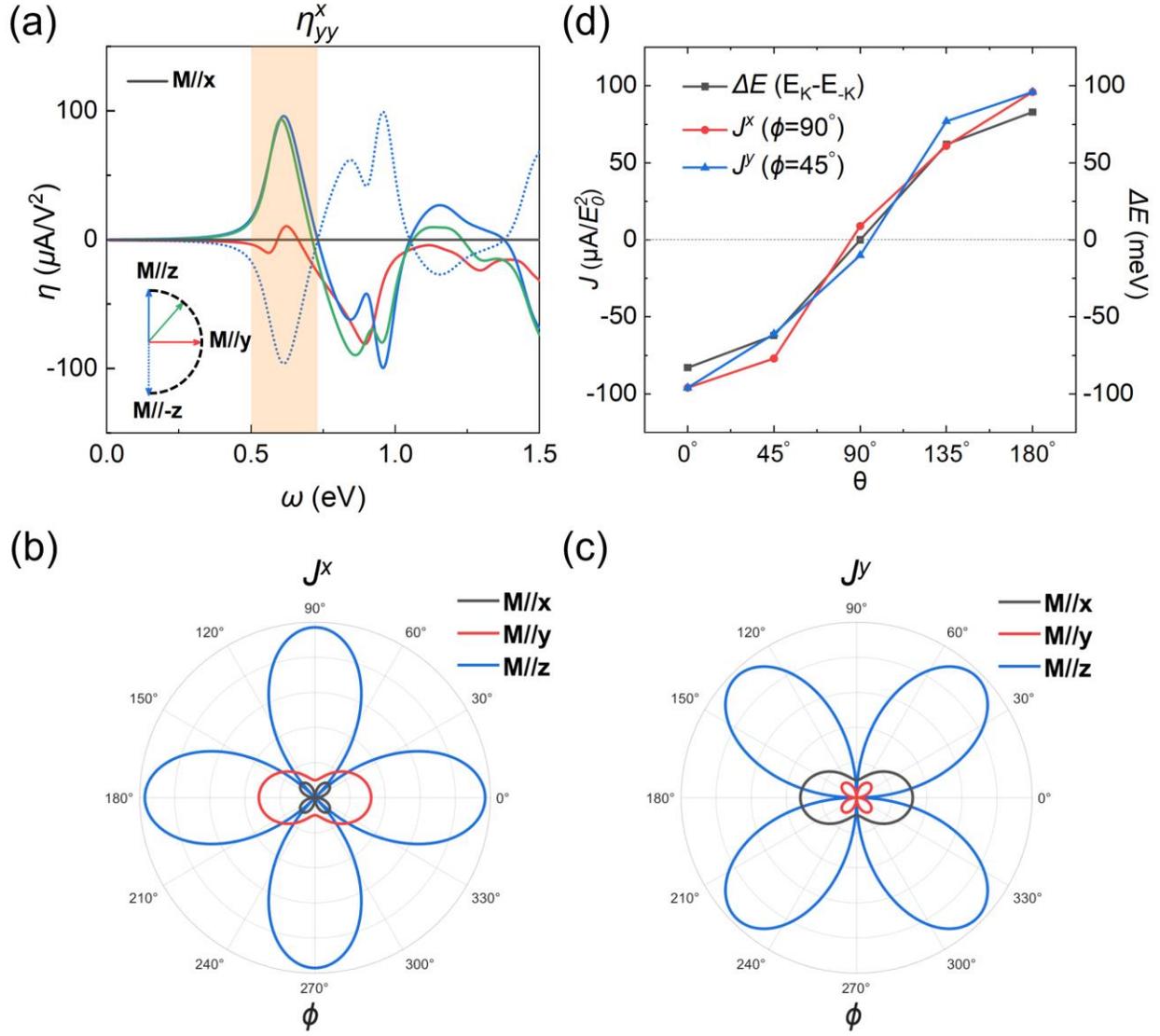

**Figure 3** Injection current in Nb$_3$I$_8$ monolayer with varying magnetic orders. (a) Injection current photoconductivity of $\eta^x_{yy}$ component with varying magnetic orders. The shaded region presents the peak enhanced by the out-of-plane magnetic order. All components reverse as spin flips from **M** ∥ **z** to **M** ∥ −**z**. (b)-(c) Angle-dependent injection current $(J^x, J^y)$ at photon energy of 0.6 eV. Angular coordinate $\phi$ denotes the angle between the *x*-axis and incident light polarization direction, and radial coordinate stands for amplitude of photocurrents. The maximum amplitude of injection current for **M** ∥ **z** is significantly greater than that for **M** ∥ **x** and **M** ∥ **y**. (d) Injection current $(J^x, J^y)$ and energy difference between $k = K$ and $k = -K$ points for the bottom conduction band as a function of $\theta$, where $\theta$ denotes the polar angle between the out-of-plane *z*-axis and spin orientation. The current is in unit of $\mu A/E_0^2$ where $E_0$ denotes amplitude of the incident light field.

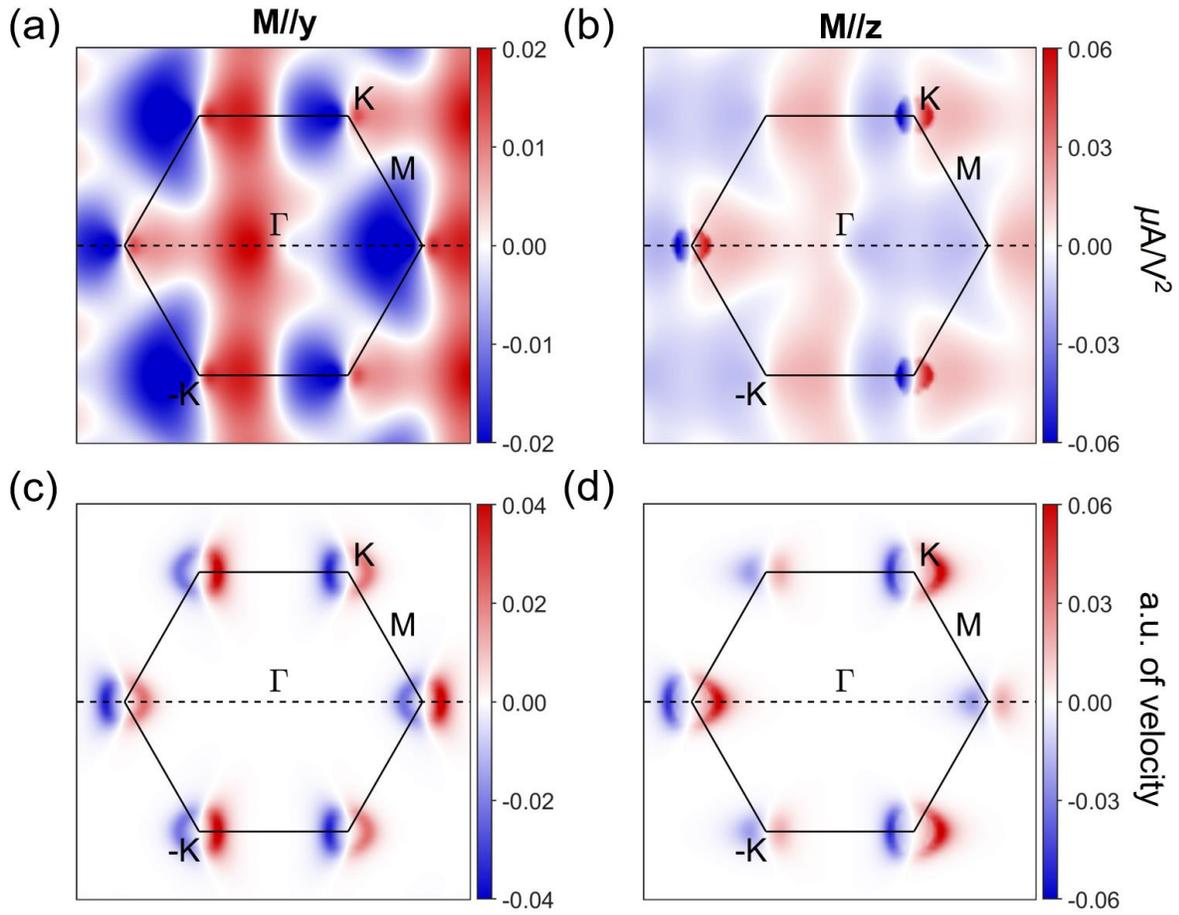

**Figure 4** (a)-(b) Distributions of group velocity difference $\Delta^x_{mn}$ between the top valance band and bottom conduction band in momentum space with **M** ∥ **y** and **M** ∥ **z**, respectively. (c)-(d) Distributions of quantum metric dipole tensor $\Delta^x_{mn}|r^y_{mn}|^2$ at photon energy of 0.6 eV in momentum space with **M** ∥ **y** and **M** ∥ **z**, respectively.

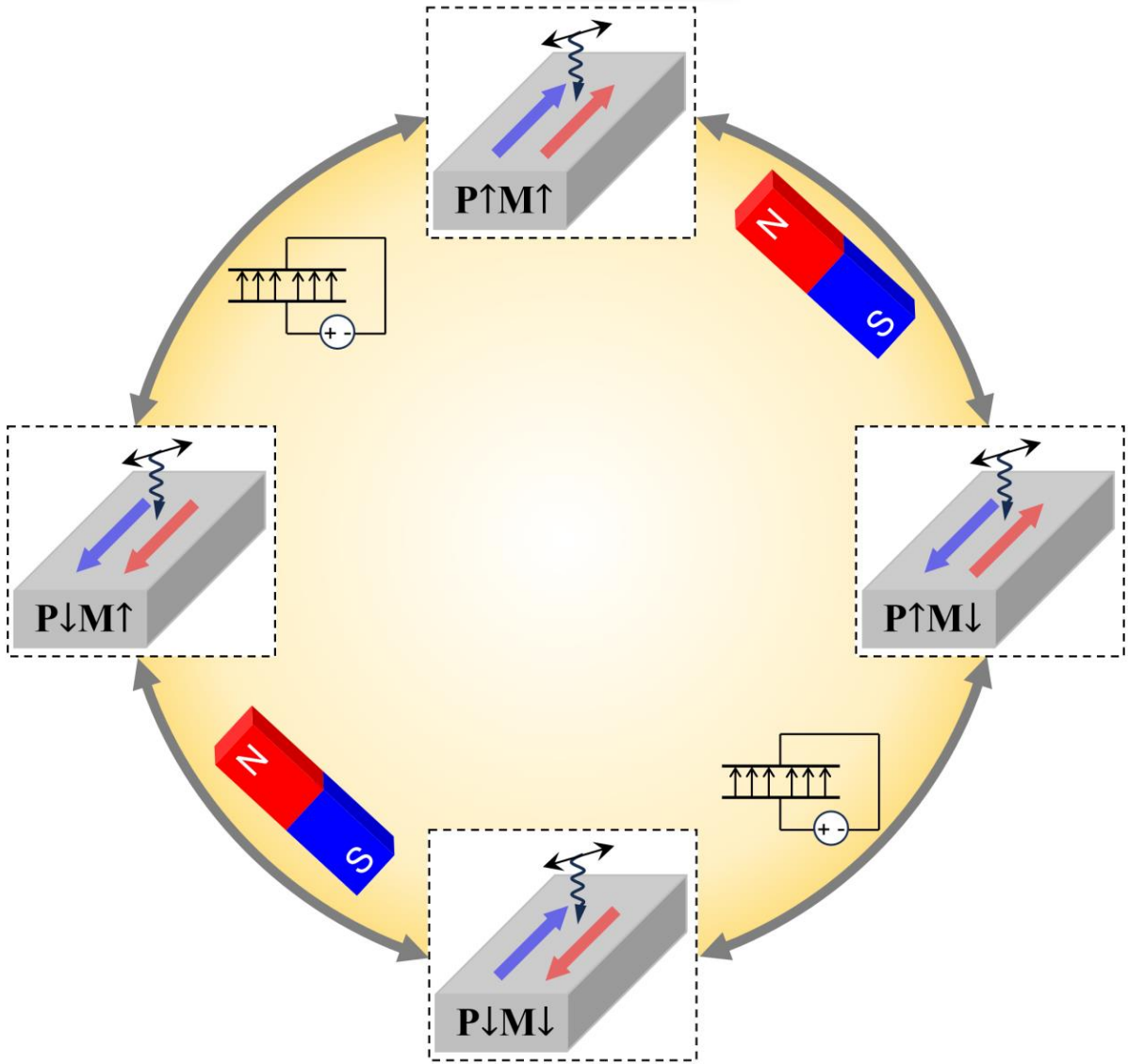

**Figure 5** Schematic plot of multiferroic phases and corresponding nonlinear photocurrent in breathing kagome lattice $Nb_3I_8$ monolayer. The four distinct phases are denoted as **P↑M↑** (top), **P↑M↓** (right), **P↓M↓** (bottom), and **P↓M↑** (left), respectively. The four multiferroic phases can be switched by applying an external electric field/magnetic field, where the injection current (blue arrow) and shift current (red arrow) are independently reversed.

**Table I** Properties of shift and injection current for typical two-dimensional multiferroic materials. The symbols, ✓ and ✗, denotes whether the photocurrent will reverse or retain its direction upon ferroelectric (FE) or ferromagnetic (FM) switching. The final column indicates the number of different nonlinear photocurrent values corresponding to FE/FM switching.

| | Symmetry operation for FE switching | NLO Response | Parity under FE switching | Parity under FM switching | # of values |
|---|---|---|---|---|---|
| Multiferroic sliding systems [13] | $M_z\mathcal{T}$ | Shift<br>Injection | ✗<br>✓ | ✗<br>✓ | 2 |
| Multiferroic breathing kagome lattice | $\mathcal{P}$ | Shift<br>Injection | ✓<br>✓ | ✗<br>✓ | 4 |

# TOC graphic

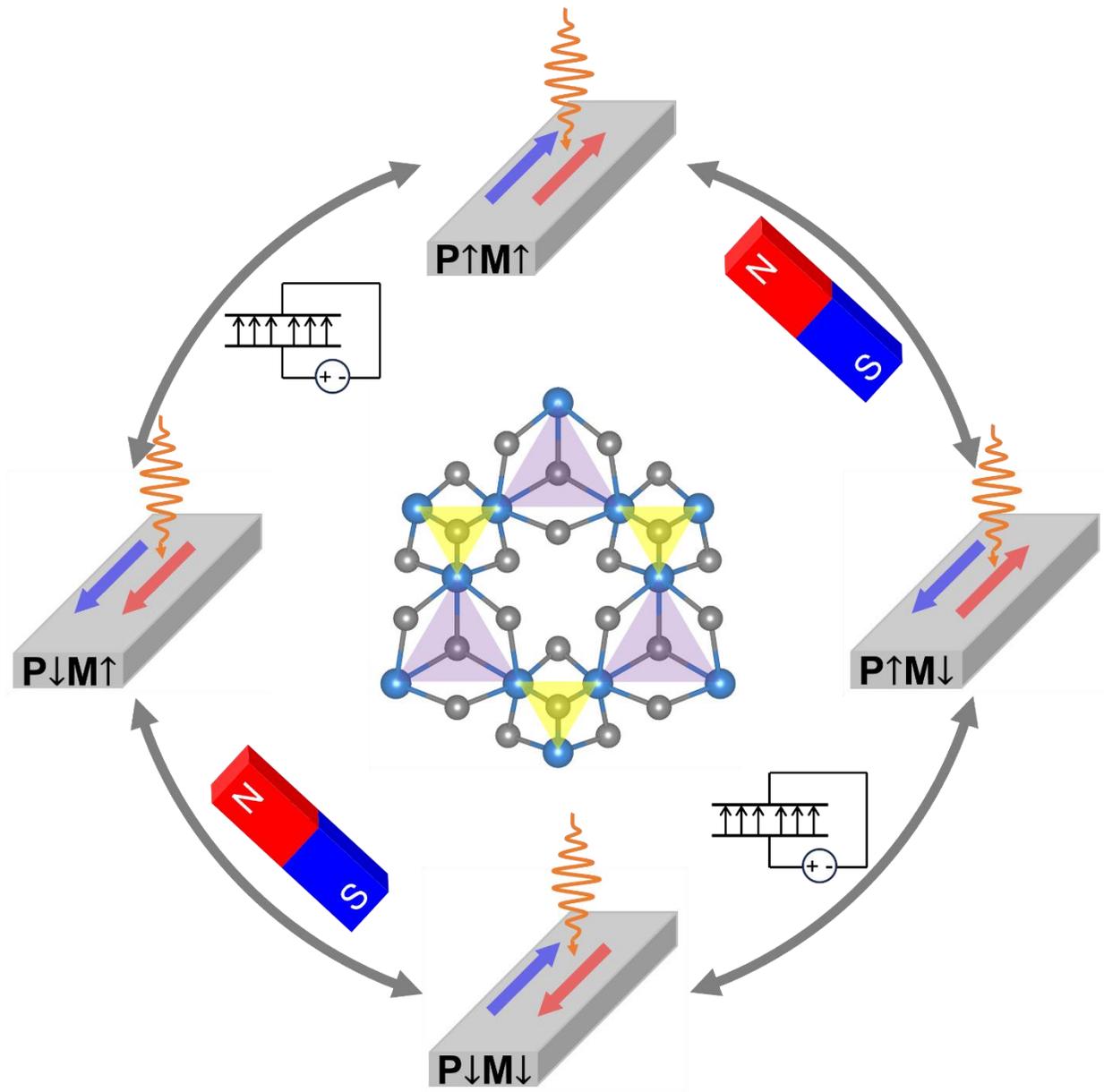